\documentclass[aps,prc,preprint,groupedaddress,floatfix,showpacs,twoside]{revtex4}
\usepackage{graphicx}
\usepackage{dcolumn}
\usepackage{bm}
\begin{document}

\setcounter{page}{1}
\pagestyle{myheadings} \oddsidemargin
4mm\evensidemargin -4mm

\begin{center}
{\bf Common Signatures of Statistical Coulomb Fragmentation of Highly Excited Nuclei and Phase Transitions in Confined Microcanonical Systems\\}
\bigskip
\bigskip
\bigskip

{J. T\~oke, M. J. Quinlan, I. Pawe\/lczak, and W. U. Schr\"oder\\}{\it Departments of
Chemistry and Physics \\
University of Rochester, Rochester, New York 14627}

\end{center}
\bigskip
\bigskip
\centerline{ABSTRACT}
\begin{quotation}
Characteristic signatures of statistical Coulomb fragmentation of highly excited nuclear systems were analyzed. It was found that in some important aspects, they coincide with perceived signatures of phase transitions in confined hypothetical pseudo-microcanonical systems, thus potentially giving rise to a false interpretation of experimental observations in terms of phase transitions. It is demonstrated that the heat capacity as derived based on experimental observations may show domains of faux negative heat capacity for the same fundamental reason a faux negative heat capacity appears in constrained numerical modeling of phase transitions in excited nuclear matter, the reason being an effective truncation of the microcanonical phase space. Similarly, selected experimental data may exhibit bimodality apparently in accordance with the truncated pseudo-microcanonical (but not the true microcanonical) calculations for confined systems.
\end{quotation}

\newpage

\section{Introduction}

For more than 2 decades now, the observed process of emission of
multiple intermediate-mass fragments commonly named
multifragmentation has inspired theoretical speculations regarding
its mechanism. Since much of the observed yield exhibits features
commonly attributed to statistical production mechanisms and since
the standard equilibrium-statistical decay codes \cite{pace,gemini}
were not able to predict any noticeable yield of intermediate-mass
fragments (IMFs) heavier than lithium, a number of models
\cite{eesm,smm,mmmc,fisher,lattice,percolation}, have been developed
to effectively parameterize some of the salient trends in the
observed IMF yields. On the other hand, because the observed IMF
yields appeared to exhibit patterns arguably expected for phase
transitions, the most prominent of these models
\cite{smm,mmmc,lattice} consider nuclear multifragmentation to be a
form of a phase transition in excited nuclear matter.  confined in a
hypothetical box.

Recently \cite{toke_surfentr}, it was shown that, when due allowance
is made for the surface entropy and thermal expansion of nuclear
matter, the basic scenario of asymmetric fission is sufficient to
explain the observed IMF yields. It is also sufficient to explain
the multifragmentation as a form of fission but generalized to
multifragment saddle shapes, all without recourse to the notion of a
phase transition. According to this fission-like scenario, the
excited system undergoes shape fluctuations with amplitudes
increasing as the surface tension diminishes with increasing
excitation energy. As the system reaches randomly a binary or
multifragment saddle configuration it is pulled apart toward
scission  by Coulomb and/or centrifugal forces. For the purpose of
the following discussion, this process that is named Coulomb
fragmentation. In fact, this is the very scenario considered by the
statistical decay code Gemini \cite{gemini}, which, however, fails
because of its inadequate account of thermal expansion of nuclei and
of the role surface entropy plays in the process. It was also shown
recently \cite{toke_unified} that binary (asymmetric) Coulomb
fragmentation is quantitatively described by equations that are
equivalent to the parameterization used by the Nuclear Fisher's
Droplet Model \cite{elliott-fisher}, which was found to fit a large
volume of experimental data. In addition, multiple Coulomb
fragmentation is consistent with the numerical procedures used in
the statistical multifragmentation codes SMM \cite{smm} and MMMC
\cite{mmmc}. The present study shows for the first time that
generalized fission or Coulomb fragmentation may share some
prominent signatures with second-order phase-transitions of
spatially confined systems and can thus be erroneously associated
with such transitions.

\section{Modeling of Second Order Phase Transitions in Confined
Pseudo-Microcanonical Systems}

For the purpose of this study, we consider a schematic model
\cite{toke_surface} that emulates essentials of
pseudo-microcanonical models SMM \cite{smm} and MMMC \cite{mmmc}, as
far as phase transitions are concerned, but at the same time allows
one to model Coulomb fragmentation \cite{toke_unified}. The model considers finite amount of neutral Fermi matter that is allowed to assume two different spatial configurations - those of a spherical mono-nucleus and of two touching spheres of equal sizes. Importantly, this model accounts for the presence of a diffuse surface in that it includes in the calculations both, the surface energy  and the surface entropy \cite{toke_surfentr} associated with this domain. The presence of this diffuse surface domain has profound effects on the the way the system evolves with increasing excitation energy, i.e., on the possible phase transition (when the system is considered externally confined) or on the onset of Coulomb fragmentation (when the di-nuclear configuration is identified with a fragmentation saddle configuration) .
In the present study, the term ''phase'' is used to describe a
macroscopically distinct state of the system that can be assigned a
definite value of a properly chosen order parameter or a vector
of order parameters. For the breakup-type of phase transitions
appearing in pseudo-microcanonical SMM \cite{smm} and MMMC \cite{mmmc} calculations, a natural choice for the ``order vector'' is the set of mass and atomic numbers of the fragments. The term second-order phase transition is used here to describe a situation where the system changes the
most likely phase it will be found in. Note that, while in the
thermodynamical limit the system such as considered here would always reside in a pure, single-phase state, a finite system will always reside in a mixed-phase state. The probability of finding it in a particular pure phase is given by the weight function proper for the kind of ensemble considered (e.g., microcanonical, grand canonical, canonical, isobaric-isothermal, etc.). Note that the above definition of phases and phase transitions is is fully consistent with conventional thermodynamics. It is reiterated here only to enhance the clarity of the chain of arguments used further below. Note also that we use here the prefix ``pseudo''in the qualification of SMM \cite{smm} and MMMC \cite{mmmc} models, in order to stress the important fact that only a truncated phase space is numerically manageable in these, as in any other calculations of systems at excitation energies of interest here. As shown further below, an incomplete accounting of
phase space may have non-trivial, qualitative consequences in the
domain of second-order phase transitions. In fact, omission of certain parts of phase space may be responsible for faux signatures of such transitions, such as apparent phenomena of bimodality and negative heat capacity reported in the literature.

With the above definition of phases, a second-order phase transition
occurs when the weight functions for different phases intersect as the value of the controlling parameter (such as, e.g., total energy or temperature) is varied. This point is illustrated using the pseudo-microcanonical schematic model of Ref.~\cite{toke_surface}, which permits just two distinct macroscopic states of an excited nuclear system - the mononuclear and a symmetric di-nuclear configuration. Obviously this is the minimum number of allowed macroscopic states or phases needed for the concept of
a phase transition to be meaningful. In this case the pseudo-microcanonical weight functions $w(E)$ for the two configurations are
functions of total energy $E$ of the system.  They are related to
conditional entropies $S_{m}$ and $S_{d}$ associated with mono- and
di-nuclear configurations, respectively, i.e., the hypothetical entropies of the system forced into either a mono-nuclear (subscript $m$) or a di-nuclear (subscript $d$ ) configuration,

\begin{equation}
w_{m/d}(N,E,V)=e^{S_{m/d}(N,E,V)}.
\label{eq:weight}
\end{equation}

\noindent Here, $N$ is the number of particles in the system and $V$ is the system volume considered here constant. Because of the above simple
microcanonical relationship between a weight function and the
corresponding conditional entropy, a (second-order) phase transition will occur at a system energy $E$, where the
conditional entropy functions for the two configurations or phases
cross, i.e., where $S_m(N,E,V)=S_d(N,E,V)$. Note that for canonical ensembles at constant particle number $N$,
temperature $T$, and volume $V$, the proper weight function can be
expressed in terms of the Helmholtz free energy $A$,

\begin{equation}
w_{m/d}(N,T,V)=e^{-A_{m/d}(N,T,V)/T}.
\label{eq:weight_helmholtz}
\end{equation}

For an isothermal-isobaric ensemble at constant $N$, temperature $T$ and
pressure $p$, the weight function is properly expressed in terms of the Gibbs free energy $G$

\begin{equation}
w_{m/d}(N,T,p)=e^{-G_{m/d}(N,T,p)/T}.
\label{eq:weight_gibbs}
\end{equation}

Obviously, for all three ensembles considered above, the crossing of
the corresponding weight functions will occur at that value of the relevant
argument ($E$ for microcanonical, otherwise $T$) for which a
crossing of the respective special thermodynamic state functions $S$,
$A$, and $G$ occurs for the two nuclear configurations or phases considered. In the thermodynamical limit, the crossing of
the relevant special thermodynamic state functions results in a
discontinuity in the respective first derivatives of these functions
which in certain mathematical representations appear as zeroes of
the partition function (e.g., on the complex plane of canonical
temperature) \cite{chomaz_iwm}. For finite systems, statistical
fluctuations cause ``spreading'' of the phase transition over a
domain of argument values resulting in the disappearance of these
singularities. Again, in terms of complex calculus, such a
disappearance can then be viewed as a result of moving of the zeroes
of the partition function on the complex argument plane to a
suitable locations. While one must acknowledge the beauty of complex
calculus, taken alone, such purely mathematical constructs have the
potential of obscuring or masking the physics. The physics of the
phase transition phenomenon appears clearer in the representation
used here, i.e., based on the notion of the crossing of relevant
weight functions and mixing of macroscopic configurations (phases)
as a result of fluctuations.

Using the Ferm-gas model for the level densities of the constituent nuclei, the conditioal entropies for the pseudo-microcanonical mono- and symmetric di-nuclear configurations can be written as

\begin{equation}
S_m(A,E^*)=2\sqrt{a_mE^*}
\label{eq:S_mono}
\end{equation}

\noindent and

\begin{equation}
S_d=2\sqrt{a_d[E^*-(E_d^{pot}-E_m^{pot})]}.
\label{eq:S_di}
\end{equation}

In Eqs.~\ref{eq:S_mono} and \ref{eq:S_di}, $a_m$ and $a_d$ are level
density parameters ({\it little-a}) for mono- and di-nuclear configurations, respectively, and $E_m^{pot}$ and $E_d^{pot}$ are the potential energies of these configurations. The level density parameters for a realistic nuclear matter distribution with diffuse surface domain can be calculated using Thomas-Fermi approximation. They can be expressed approximately in terms of volume and surface contributions as \cite{toke-swiatecki}

\begin{equation}
a_m={A \over 14}(1+4A^{-1/3}) MeV^{-1}
\label{eq:littlea}
\end{equation}

\noindent and

\begin{equation}
a_d=2{A \over 28}[1+({A\over 2})^{-1/3}] MeV^{-1}.
\label{eq:littlea}
\end{equation}

The potential energies $E_d$ and $E_m$ for di- and mono-nuclear
configurations can be calculated from the liquid drop model
\cite{mye69} such that their difference is equal to the difference
in surface energies for the two geometries

\begin{equation}
E_d-E_m=c_{Surf}(F_2^d-1)A^{2/3}.
\label{eq:surfenergy}
\end{equation}

Here $c_{Surf}$ is the surface energy coefficient, and
$F_2^d=2^{1/3}$ is the ratio of the surface area of the symmetric di-nucleus to that of a sphere of the same volume.

The presence of a surface specific contribution to the level density
parameter is essential for the present study. In particular, it is crucial for a quantitative understanding of the Coulomb fragmentation, both of binary and multiple fragmentation. This surface term allows one to approximately account for the excess entropy per nucleon (surface
entropy) contributed by the diluted matter in the diffuse surface
domain as compared to that of a nuclear system with homogeneous matter distribution. It is the presence of this extra surface entropy for deformed shapes that at elevated excitation energies greatly enhances the system chances for arriving at deformed saddle configurations and thus to undergo Coulomb fragmentation. It is also the presence of surface entropy that makes weight functions for different fragmentation phases to intersect at characteristic excitations \cite{toke_surface} leading to second-order phase transitions. This feature is illustrated in the figures further below.

It is important to note that the present schematic model differs conceptually from the statistical multifragmentation models SMM \cite{smm} and MMMC \cite{mmmc}, but it relies on the same basic mechanism of phase transition as the latter models \cite{toke_unified}. And so, what is in the present model achieved by an explicit modeling of the diffuse nuclear surface domain, is achieved in SMM
\cite{smm} and MMMC \cite{mmmc} models by extra entropy associated with the implicit random motion (``rattling'') of spherical fragments within a hypothetical confinement (named arbitrarily as ``freezeout'' volume) of a volume fine-tuned to obtain a desired fit to select experimental observations. Note that such rattling of the spherical fragments leads to an effective spreading of the time-averaged matter distributions of individual fragments featuring diffuse surface domain. The extra ``rattling'' entropy is then to a good extent equivalent to entropy generated in this time-averaged diffuse surface domain. However, unlike SMM and MMMC recipes relying on an $ad hoc$ use of hypothetical ``oversize'' confinement vessels and on a lengthy statistical process of nucleon evaporation and re-synthesis into fragments, the present model derives the surface entropy consistently from the well understood liquid-drop and Fermi gas models of excited nuclei.

\begin{figure}
\includegraphics[width=12cm]{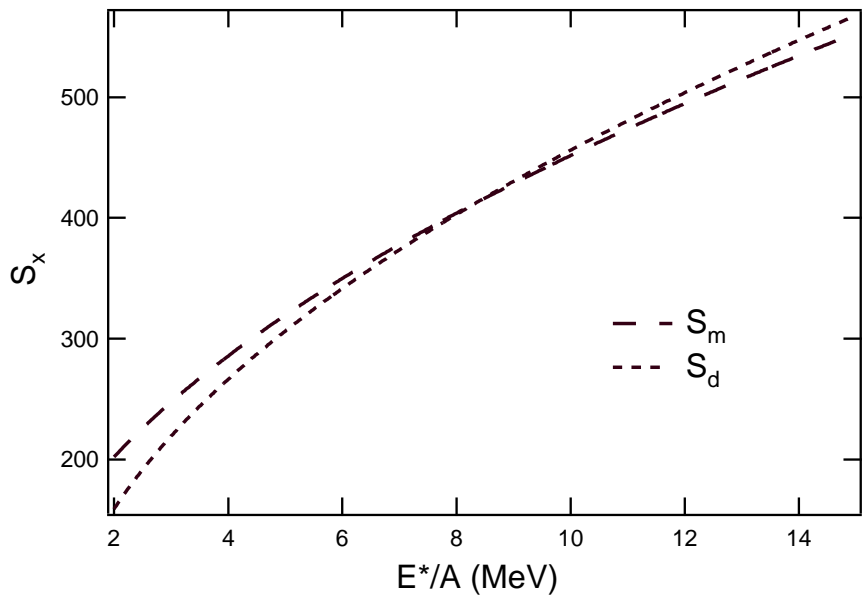}
\caption{Functional dependence of conditional entropies (top panel) and relative population probabilities (bottom panel) for mono- and di-nuclear phases, $S_m$ and $S_d$, respectively, on excitation energy per nucleon exhibiting a crossing at approx. 8.35 MeV/u.}
\label{fig:crossing}
\end{figure}

\begin{figure}
\includegraphics[width=12cm]{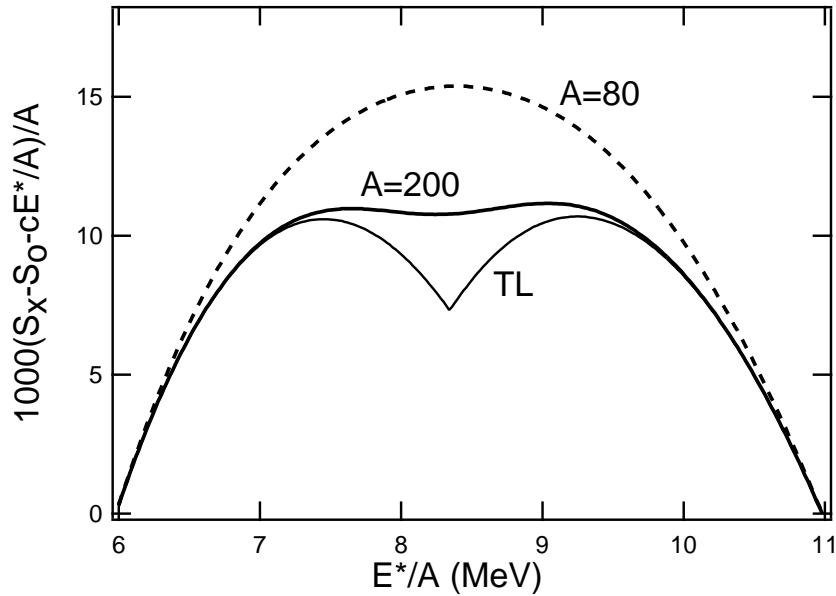}
\caption{Reduced entropy vs. excitation energy per nucleon for the two-phase system of mono- and
di-nuclear configurations and three sizes of the system - thermodynamic
limit (TL), ``large'' with A=200, and ``small'' with A=80.(See text)} \label{fig:convex}
\end{figure}

Results of model calculations for a two-phase system, allowing mono-
and di-nuclear configurations within the schematic formalism
\cite{toke_surface} reiterated above, are shown in Figs.~\ref{fig:crossing} -
\ref{fig:convex}. Fig.~\ref{fig:crossing} illustrates the intersecting of the two ($m$ and $d$) conditional entropy functions at an ``cross-over'' excitation energy of $E/A\approx 8 MeV/u$, which also signifies crossing of the corresponding weight functions for the two configurations. Since at low excitations $S_m>S_d$, the system will dominantly reside in the mono-nuclear state or phase. As the excitation energy increases approaching the crossing point, the system will more frequently fluctuate away from the mono-nuclear configuration into the di-nuclear domain. In the vicinity of the ``cross-over'' energy the two phases will coexist, but not in the sense in which macroscopic amounts of liquid and gas coexist in first-order phase transitions. Rather, at any moment in time the system as a whole is either in one or the other of the two pure macroscopic states. At higher excitation energies, beyond the cross-over point, the system will be dominantly in the di-nuclear state. Note, that the crossing of the weight functions or conditional entropy functions represents a second-order phase transition as there is no latent heat transferred in the transition. The thermodynamics of the second-order phase transition can be studied by exploring the evolution of the entropy of the system with increasing excitation energy. The entropy is given by the logarithm of the microcanonical partition sum

\begin{equation}
S(E)=ln[e^{S_m(E)}+e^{S_d(E)}],
\label{eq:entropy}
\end{equation}

\noindent where $S_m(E)$ and $S_d(E)$ are conditional entropies for
the mono- and di-nuclear configurations.

The reduced entropy function for the ensemble considered here is illustrated in Fig. 2, for three different sizes of the system - infinite ($TL, A->\infty$), large ($A=200$), and small ($A=80$). The reduced entropy is obtained from model entropy $S_{[m,d]}$ by subtracting a linear function such that the reduced entropy is zero at the boundaries of the display region. Additionally, the reduced entropy was rescaled by a factor 1000/$A$. As seen in Fig.~\ref{fig:convex}, in thermodynamic limit the reduced entropy (and hence the model entropy)  features a kink, but always remains a concave function of energy. Moderate fluctuations present in a finite system of A=200 turn the kink into a convex domain or ``convex intruder'', which is then ``healed'' by larger fluctuations in an A=80 system. Note that a subtraction of a linear function does not affect the character of the function as far as convexivity is concerned.

\begin{figure}
\includegraphics[width=12cm]{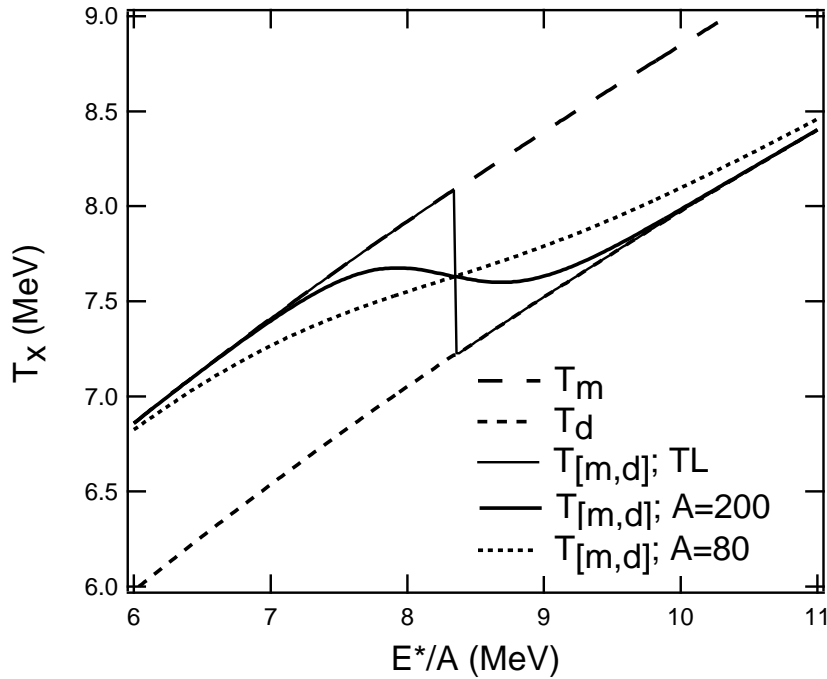}
\caption{Caloric curves for the two-phase system of mono- and
di-nuclear configurations for three sizes of system - thermodynamic
limit (TL), ``large'' with A=200, and ``small'' with A=80. Also
shown are caloric curves for pure mono- and di-nuclear ensembles,
labeled $T_m$ and $T_d$, respectively.} \label{fig:caloric}
\end{figure}

The convex intruder in the functional dependence of entropy on energy shows up in a caloric curve as a domain of negative heat capacity. This is illustrated in Fig.~\ref{fig:caloric}, where the apparent thermodynamic temperatures ($T_x$) are displayed for the three cases of system sizes considered. The thermodynamic temperature plotted in Fig.~\ref{fig:caloric} is calculated
according to the standard microcanonical expression for the average
temperature

\begin{equation}
T^{-1}=\beta={dS\over dE}
\label{eq:temperature}
\end{equation}

As seen in Fig.~\ref{fig:caloric}, in the thermodynamic limit there is a singularity in the temperature function forming a discontinuous jump. This singularity is not present for finite systems in which mixing of configurations by fluctuations render the entropy function $S$ differentiable at any energy. However, in finite systems there is still a telltale remnant of the crossing in the form of a noticeable departure of the average temperature from monotonic behavior as a function of energy. For larger systems, where the fluctuations are weaker, this irregularity may take the appearance of negative heat capacity, as seen in Fig.~\ref{fig:caloric} for the case of A=200.

\subsection{Apparent Negative Heat Capacity}

The presence of irregularity or non-monotonicity in the caloric curve of $T(E)$ appears possible in the phase transition domain of physical systems. However, the presence of negative heat capacity in such a domain, such as seen in Fig.~\ref{fig:caloric} is difficult, if not impossible, to prove based on numerical modeling. This is so because it is technically impossible to include in the model calculations all energetically allowed microstates of a physical system and thus evaluate the true entropy function and the true microcanonical temperature. Out of necessity, one always deals numerically only with subspaces of the untreatably large true microcanonical phase space. Therefore, only apparent or pseudo thermodynamic potentials
and functions are evaluated and not the true ones. It is rather obvious that ``$a priori$'' it is not possible to tell whether any particular fine (as opposed to gross or average) trend in an apparent thermodynamical quantity in a model system reflects its true thermodynamical counterpart in a physical system. In other
words, as a matter of principle, it is impossible to predict fine
trends in thermodynamical quantities of physical systems based on truncated (incomplete phase space) numerical calculations such as used by SMM \cite{smm}, MMMC \cite{mmmc}, and lattice gas models \cite{lattice}. In
particular, as far as the claims of negative heat capacity are
concerned, it is demonstrated further below that an apparent
negative heat capacity may result trivially from excluding
microstates associated with macrostates intermediate between the
ones arbitrarily identified as separate phases.

\begin{figure}
\includegraphics[width=12cm]{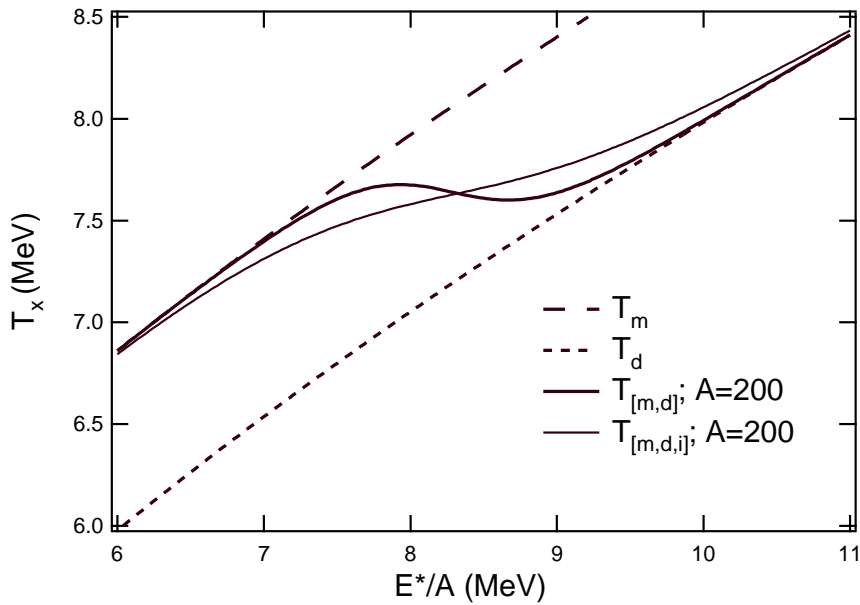}
\caption{The ``healing'' effect of an inter-phase configuration on the apparent caloric curve for the two-phase system of mono- and
di-nuclear configurations for a A=200 system.} \label{fig:healing}
\end{figure}

One notes first that physically it is impossible to have a nuclear system allowing only two distinct macroscopic configurations similar to the one used in this study. In fact, it is impossible to have a physical system allowing any number of discrete macroscopic configurations but not allowing continuous paths connecting these configurations. Here, it is impossible for a spherical mono-nucleus to transition to symmetric di-nuclear configuration without passing through a continuous sequence of intermediate states of various intermediate deformations. The same is true with respect to the configuration spaces considered by SMM \cite{smm}, MMMC \cite{mmmc}, and lattice gas models \cite{lattice}. Importantly, the intermediate macroscopic configurations have a noticeable influence on the apparent trends in thermodynamic quantities exactly in the domain of the phase transition, i.e., where the weight functions for the phases of interest cross. This is so, because the weight functions of the intermediate configurations are guaranteed to cross with the weight functions for the phases of interest in the immediate vicinity of the crossing of the latter and, hence, make a noticeable contribution to the overall partition function selectively only in this domain. Exclusion of these intermediate configurations depletes the model entropy locally (on the energy scale) and may give rise to a ``convex intruder'' in the entropy function $S(E)$. Such an apparent or $faux$ convex intruder will propagate to other representations of apparent thermodynamic observables and, in particular, will result in a caloric curve featuring a  domain of $faux$ negative heat capacity. This rather trivial mechanism of generating $faux$ negative heat capacity in truncated-space microcanonical calculations is illustrated in more detail in Fig.~\ref{fig:healing}.

\begin{figure}
\includegraphics[width=12cm]{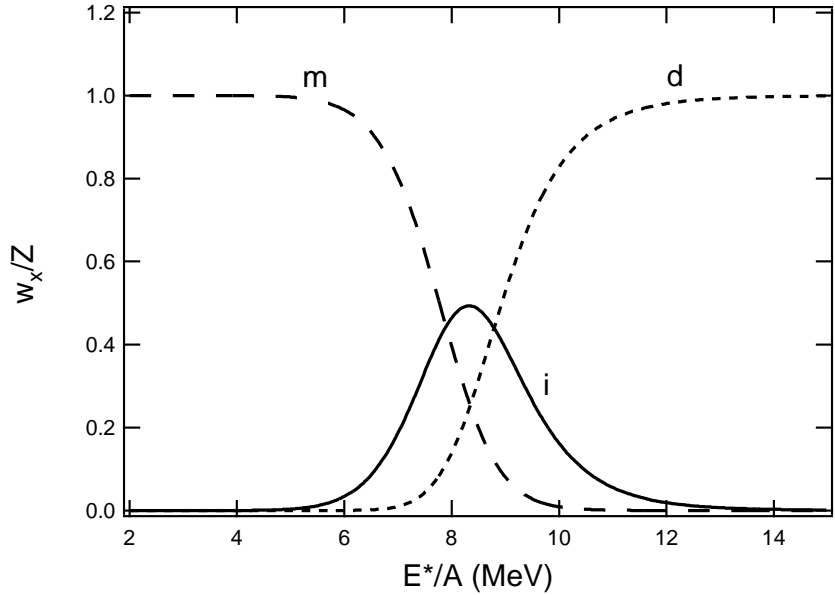}
\caption{Normalized weight functions for three macroscopic configurations, mono- and di-nuclear (subscripts $m$ and $d$)and intermediate ($i$) as functions of energy.} \label{fig:transition}
\end{figure}

\begin{figure}
\includegraphics[width=12cm]{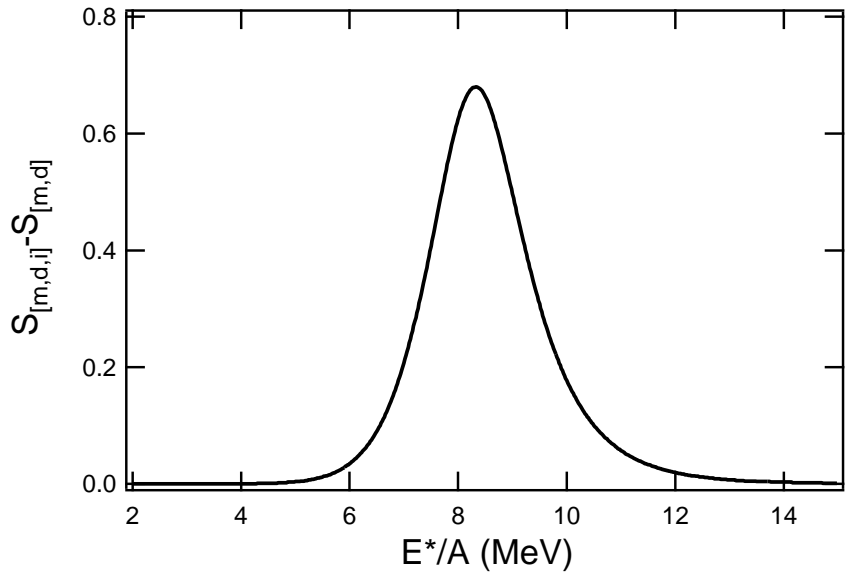}
\caption{The additional entropy associated with the inter-phase configuration with $F_2^i=0.5(F_2^m+F_2^d)$ as a function of excitation energy.} \label{fig:deltas}
\end{figure}

Fig.~\ref{fig:healing} illustrates effects of the exclusion of classes of microstates associated with just one macroscopic configuration intermediate between the two ``base'' ones associated with mono- and di-nuclear phases. The inter-phase configuration has here a deformation parameter of $F_2=1.13$, i.e. a value half-way between the respective values for mono-nuclear ($F_2=1$) and di-nuclear ($F_2=2^{1/3}$) configurations.
As seen already in Fig.~\ref{fig:convex}, the apparent entropy $S_{[m,d]}(E)$ for a system including only the two base configurations ($m$ and $d$) exhibits a convex intruder in the vicinity of the crossing point of the conditional entropy functions $S_m(E)$ and $S_d(E)$. The caloric curve $T_{[m,d]}(E)$ deduced for a hypothetical system with such a drastically truncated phase space features negative heat capacity in the vicinity of this crossing  (see Fig.~\ref{fig:caloric}. As seen now in Fig.~\ref{fig:healing}, already a ``casual'' addition of just one intermediate inter-phase ($i$) configuration with the deformation parameter $F_2^i=0.5(F_2^m+F_2^d)$ changes the apparent caloric curve
$T_{[m,d,i]}(E)$ qualitatively in such a way that the domain of apparent negative heat capacity is eliminated. How in the calculations the negative heat capacity vanishes can be understood from Figs.~\ref{fig:transition} and \ref{fig:deltas}. These figures explain implicitly, how an omission of relevant macro-configurations gives rise of an apparent or $faux$ negative heat capacity.

As seen in Fig.~\ref{fig:transition} the ``inter-phase'' configuration ($i$) plays a noticeable role selectively only in the vicinity of the cross-over point of the weight functions for the two phases considered. Accordingly, as seen in Fig.~\ref{fig:deltas}, it contributes to the overall apparent entropy only in the energy domain of the phase transition and not much beyond it. It is the presence of this additional entropy, not accounted for in the truncated two-phase phase space that restores the overall concavity of the entropy as a function of energy.

The present schematic calculation demonstrates that, in the
vicinity of an anticipated phase transition (crossing of weight
functions), apparent thermodynamic quantities behave qualitatively
differently when certain classes of valid microstates are excluded
from the numerical calculations. It is not possible to
guarantee that all relevant states are included in any realistic
model description. Therefore, as a matter of principle, truncated model calculations do not provide a sound foundation for claims regarding fine trends in true thermodynamical quantities, and especially so in the vicinity of phase transitions.

\subsection{Apparent Bimodality}

Recently, various suggestions have been made \cite{gross_book,chomaz_iwm} to the extent that bimodality of certain thermodynamic functions may serve as a signature of phase transitions. The perceived bimodality is then used to infer the presence of a convex intruder in the entropy as a function of excitation energy. In view of
the above discussion regarding the origins of the latter intruder, it is justified to ask whether an apparent numerical bimodality is a
reflection of a true bimodality or simply the result of an arbitrary truncation of the model phase space.

Obviously, in a schematic model that is inherently bimodal, like the
one employed here, and allowing only two macroscopic configurations,
many numerical observables are expected to exhibit a trivial
bimodality. For example, at any fixed energy, the numerical system (but not a physical one) will statistically jump from one macroscopic configuration to another one with a much different potential energy. Accordingly, the model temperature will jump from one value to other, leading to the corresponding bimodal temperature distribution. Similarly, in a canonical system, the system energy will jump from one value to the other, leading to an apparent bimodality in the energy distribution. A physical system, however, must pass through intermediate macroscopic states with potential energies intermediate between those for the two ``base'' configurations. When such intermediate configurations are included in the model calculations,
the gap between the peaks of the bimodal distributions will be
filled. To reiterate, since it is not possible to guarantee that in a model (numerical) phase space all relevant microstates are included,
generally, numerical model calculations cannot provide a sound
foundation for statements regarding the bimodality or lack thereof
of distributions of true thermodynamic observables for physical sysytems. However, in some cases, one can prove, based on thermodynamic principles that certain distributions must be mono- and not bi-modal.

A prominent case of suspected bimodality pertains to the behavior of isothermal-isobaric ensembles (fixed number of particles $N$, fixed temperature $T$, and fixed pressure $p$). For example, Ref.~\cite{gross_book} asserts that, in a ($N$,$T$,$p$) model of a cup of coffee at the phase transition temperature, the coffee will be in the state of pure liquid 50\% of time and in the remaining 50\% of time, it will be in the state of a pure vapor. Such narration and the
associated model calculations neglect all macroscopic configurations where part of coffee is in liquid and part is in gaseous state. Evidently, it is not possible for an entire physical system to jump from one volume (liquid) to another (gas) without passing through a whole sequence of intermediate volumes. Such jump would entail instantaneous transfers and equilibration of finite amounts of (latent) heat between the system and the isobaric thermostat, and an instantaneous expansion or contraction of matter involving velocity fields of infinite magnitude.
The perceived bimodality is in this case simply the result of an incomplete conceptual and numerical modeling, i.e., neglect to account for all states where liquid and gas coexist. Accordingly, the deduced caloric behavior of the system exhibits only a $faux$, but not physical negative heat capacity. Interestingly, in this case one can prove based on thermodynamical principles, not contested in Refs.~\cite{chomaz_iwm,gross_book,gross_iwm}, that inclusion of states of coexisting liquid and gas must result indeed in a mono-modal distribution. This is so, because the weight functions for different macroscopic configurations of an ($N$,$T$,$p$) ensemble are expressed via Gibbs free energy $G$ as seen in Eq.~\ref{eq:weight_gibbs} and because at fixed $T$ and $p$ this energy does not depend on volume. Therefore, any intermediate volume of the system between the volumes associated with pure phases, $V_{liquid}$ and $V_{gas}$, must be equally probable, leading at phase coexistence (infinitely narrow domain in $p$ for a given temperature $T$ of the thermostat) to a ``flat-top'' mono-modal
distribution in $V$ spanning uniformly a domain between $V_{liquid}$
and $V_{gas}$. Unlike a bimodal distribution, such a ``flat-top''
distribution has a simple physical interpretation in which the
system has no preference regarding the relative number of particles
in the two phases. The system will simply randomly fluctuate between
the states with various numbers, e.g., of gas particles,
deriving/returning energy from/to the heat bath as particles are
transferred between liquid and gas. Such a ``random-walk'', not
subject to any driving force other than the compressional at the
natural boundaries of the domain (pure liquid at one end and pure
gas at the other), leads to a uniform distribution in system
volumes. Note, that the ``flat top'' of the Gibbs free energy distribution corresponds to the Maxwell plateau in the isotherm plotted in a pressure $p$ $versus$ volume $V$ representation.

\subsection{First Order Phase Transitions in Small Truncated Systems}

Quite generally, first order phase transitions in small confined
microcanonical systems can be viewed as a ``rapid succession'', along
the energy axis, of crossings of weight functions for
configurations with, e.g., successively increasing number of gas
particles. In the absence of other macroscopic configurations, such
a succession of second-order phase transitions will reveal
characteristics of a first order phase transitions. In this case, the latent heat of evaporating a single nucleon is the amount of energy
needed to pass from one crossing point to the next one with the
second-order transition occurring without infusion of energy. In
nuclear systems, the latent heat for the emission of individual
consecutive nucleons may fluctuate significantly, e.g., due to the
pairing energy. Because of this uncertainty in latent heat for
consecutive acts of evaporation, there is no advantage from
parameterizing the observations in terms of the first-order phase
transitions, with the scenario of successive second-order phase
transitions offering a better understanding of the underlying
phenomenon. One notes that in the SMM and MMMC models, the population of gaseous phase is by design suppressed (in terms of loss of entropy) by the arbitrary resetting of its interaction energy to zero at any density. At the same time, the relative weights of multifragment configurations are in these calculations artificially enhanced by the arbitrary presetting of the nuclear matter incompressibility modulus to infinity. Therefore, in these models the crossings of the weight functions for many multi-fragment configurations occur in the same energy domain where the crossings occur for liquid -- gas configurations, rendering a true first-order liquid--gas transition undetectable. It is of interest to check what the outcome of SMM and MMMC type of  calculations would be for a ``numerical'' matter emulating more closely a physical nuclear matter, i.e., having the same effective EOS as the latter. Would then the model matter still end up fragmented into various sets of IMFs with relatively few free nucleons floating in the freezeout volume or would it end up as a large thermally expanded liquid residue surrounded by a gas of interacting
nucleons. Such an interest is purely academic as it is not possible
experimentally to confine the system to a fixed volume and thus to
confirm or falsify such ``improved'' model predictions.

\section{Phase-Transition-Like Signatures in Coulomb Fragamentation}

Coulomb fragmentation is a process similar to binary fission but
generalized to arbitrary saddle shapes, including multifragment shapes
\cite{toke_unified}. The process can be described approximately in terms of microcanonical thermodynamics of an excited nuclear system confined transiently by surface tension to a finite domain of phase space. The boundary of this domain is defined by all possible transition states, including particle evaporation thresholds and fragmentation/multifragmentation saddle states.
Microcanonical thermodynamics requires the system to explore all energetically allowed microstates with equal probability so as to ensure maximum entropy. In the course of
such an ``exploration'', and before every possible microstate has
been visited, the system ``wanders'' into a dorway or transition state
that leads to an open decay channel. If this is a particle
evaporation channel, the system continues its journey toward
equilibrium under a new identity, the newly reduced number of
particles. If the transition state is a fragmentation saddle shape,
the system is driven toward scission, all the while continuing its
exploration of the accessible phase space. The basic macroscopic
phenomena associated with equilibration are thermal expansion and
thermal shape fluctuations. The latter are responsible for Coulomb
fragmentation, i.e., for bringing the system occasionally to a
particular saddle configuration from which it is dynamically driven
apart by Coulomb forces. The above scenario is identical to a compound
nuclear fission scenario and is identical to the IMF production scenario in statistical decay models. A novel observation made only recently \cite{toke_surfentr} is that of the role of the extra entropy associated with the diffuse surface domain in facilitating shape fluctuations. 

\subsection{Signatures of Second-order Phase Transitions}

The probability for an excited nuclear system to arrive at a particular saddle configuration is given by a respective weight function identical to the one discussed further above in the context of phase transitions
of confined systems. However, instead of different phases, now different saddle configurations are considered. The weight functions for various saddle shapes are still given by Eq.~\ref{eq:weight} and
intersect at characteristic excitation energies, since saddle shapes with larger surface area and larger potential energy are also associated with larger level density parameters. The former makes the macrostates with larger surface areas (larger values of the $F_2$ parameter in Eq.~\ref{eq:surfenergy}) accessible only at excitation energies higher than needed to excite more spherical configurations. The larger level density parameter makes the
conditional entropy of more deformed configurations grow faster with
increasing excitation energy than the conditional entropy of
more spherical shapes. As a result, the weight functions for different fragmentation channels may cross and leave signatures reminiscent of of second-order phase transitions. It is worth noting that the Coulomb fragmentation phenomena include intrinsic, inherent filter which allows to pass only saddle configurations and not the configurations intermediate between the deformed saddle and spherical reference configuration similar to what is done in truncated-space model calculations. However, to extract true and not apparent values of thermodynamical parameters from the experimental data on Coulomb fragmentation, one needs to include these experimentally ``invisible'' configurations in the model space. And again, it is virtually impossible to guarantee that deduced fine trends, such as negative heat capacity, are not artifacts of omission of relevant configurations in the model phase space.

In view of the fact that for reasonably-sized systems, there is a large number of possible saddle shapes and of corresponding crossings of the respective weight functions, one may wonder if and under what circumstances irregularities may still show up in the apparent trends of thermodynamic quantities inferred from cursory sampling of an unstable nuclear system via Coulomb fragmentation channels. Obviously, crossing points of the configurational weight functions will not be distributed uniformly on the energy scale. Rather, their distribution is expected to fluctuate and could even exhibit local statistical bunchings. When a particular bunching is sufficiently strong, it may possibly show up, e.g., in apparent caloric curve, as an irregularity and perhaps even simulate negative heat capacity. Such an irregularity, however, should not be taken as a signature of a phase transition or of a transition from one preferred individual fragmentation channel to
other. Rather, such a phenomenon will be a reflection of statistical fluctuations in the distribution of crossing points, reminiscent of
Ericson fluctuations \cite{ericson} in compound nucleus decay at low-excitations.

\subsection{Signatures of First-order Phase Transitions}

As was pointed out in Ref.~\cite{toke_unified}, an excited nucleus is inherently a two-phase system, consisting of liquid quasi-uniform bulk matter and diluted surface-domain matter. With increasing excitation energy, in a quest for maximum entropy, the system will transfer matter from the bulk to the surface domain in what is a true first-order liquid-gas phase transition. Interestingly, the presence of such a transition will manifest itself in the excitation function of the Coulomb fragmentation as a fast succession of crossings of weight functions for fragmentation saddle configurations with increasing surface areas. This is a true first-order liquid - gas (bulk matter liquid - surface domain gas) phase transition with a latent heat representing the amount of energy needed to transfer nuclear matter from bulk to the diluted surface domain.

\section{Summary}

It appears rather obvious that trends in apparent thermodynamical
quantities modeled in numerical simulations of nuclear systems that sample only a fraction of the microcanonical or canonical phase space do not necessarily represent corresponding trends in their actual physical counterparts. This observation poses the question whether such model trends reflect at least qualitatively the correct underlying physics and under what circumstances the numerical trends in apparent quantities may even be qualitatively misleading. The present
study demonstrates that qualitative discrepancies between physical processes and model interpretation are likely to
occur for phenomena involving phase transitions, the very focus of many numerical modeling attempts. The numerical examples discussed above show that the omission of macroscopic configurations intermediate between those associated with different phases depletes the partition function selectively exactly in the domain of interest and not much beyond it. As a result of such a depletion, one then observes a deficit in apparent entropy (so called ``convex intruder'') and an apparent or faux negative heat capacity in the model calculations. No such deficit would be present in a calculation performed with a more complete phase space. Similarly, the present study demonstrates that an apparent bimodalities calculated for observable distributions are results of an incomplete configuration space admitted in simulations.

While the modeling of a confined system is of purely academic
interest, modeling of the decay modes of excited nuclear systems is
of a practical interest. This latter venture aims at understanding experimental observations at a deeper level, rather than just fitting observations with various numerical parameterizations. The present study demonstrates, for the first time, that Coulomb fragmentation exhibits signatures deceptively similar to those of phase transitions in confined systems. Both sets of signatures are here reflections of crossings of respective weight functions with increasing system excitation. Hence, this work offers a plausible explanation why Coulomb fragmentation can easily be mistaken for a nuclear phase transition, either of first or second order. It is also important to note that in measurements of binary or multi-fragment decay, only a fraction of the complete underlying phase space is sampled, limited to saddle-point configurations, but excluding all ``intermediate'' pre-saddle configurations. Therefore, also experimentally one can determine directly only apparent thermodynamic quantities and their trends, but not the trends in the true thermodynamic quantities describing the highly excited nuclear systems. A more accurate determination of trends has to rely on a comparison with simulations admitting a representative, sufficiently large manifold of the configuration space.

Because of the large number of possible saddle configurations in
reasonably sized systems and because of thermal fluctuations, it is
unlikely that any single second-order ``phase'' transition between two distinct saddle configurations will leave a distinct and unambiguous experimental ``fingerprint''. More likely, phase-transition-like signatures could be observable in Coulomb
fragmentation when random, statistical ``bunching'' of the
crossing points of weight functions for different saddle
configurations occurs on the energy scale. Such statistical fluctuations could result in observable irregularities, e.g., on caloric curves, somewhat reminiscent of Ericson fluctuations \cite{ericson} in compound nucleus decay.

While it is not yet clear, what there is to learn from the
irregularities of apparent thermodynamic quantities extracted
from nuclear fragmentation data, it is clear that an analysis of experimental data in terms of Coulomb fragmentation (generalized fission) has important implications for our understanding of properties and behavior of the surface domain of excited nuclei. In
particular, such data may allow one to probe the nuclear EOS of diluted surface domain matter and also of the bulk nuclear matter diluted
through thermal expansion. It is rather obvious from the lessons of compound nuclear fission that the stability of nuclear systems against fragmentation depends critically on the presence of the diffuse surface domain and the properties of this domain. Also, it has been pointed out in Ref.~\cite{toke_unified} and further above, that because of the fact that matter in the surface domain has on average a density, binding energy, and level density parameter different from the corresponding bulk matter quantities, finite nuclei are inherently two-phase systems. In such two-phase systems, matter is transferred from bulk to surface domain, as excitation energy is raised, which is a true first-order liquid-gas phase transition. It is also natural to expect that at high excitation energies, the surface tension generated by the density gradient in the surface domain will vanish and give rise to critical phenomena.

The signatures of Coulomb fragmentation can be probed in greater detail experimentally by studying excitation functions for individual fragmentation channels. Since the process appears to be controlled by the manner in which weight functions for various multi-fragment saddle shapes cross in succession, the yield for individual fragmentation channels is expected to exhibit a rise and fall with increasing excitation energy. (See Fig.~\ref{fig:transition}, middle, Gaussian-like curve.) This may potentially open a completely new venue of experimental exploration of Coulomb fragmentation - spectroscopy of multifragment saddle configuration, where the excitation functions for various channels are studied, e.g., as functions of fragment sizes and isospins. Furthermore, it may be possible to use the location of the peaks in various yields on the energy scale as a measure of excitation energy, i.e., as a tool of calorimetry of highly excited nuclei. Also, there must be situations, where two or more weight functions run quasi-parallel avoiding each other, but crossing some other weight functions. This effect will give rise to gentle Boltzman-like scalings for some relative yields embedded in phase-transition like scalings of some other yields. This situation may arise when two saddle shapes differ in isospin but not so much in surface area, leading to iso-scaling.

With the above experimental opportunities in mind, a more coordinated theoretical effort appears warranted aiming to achieve a better
quantitative understanding of the evolution of the properties of the surface domain with excitation energy and other variables. To understand nuclear dynamics on a deeper level, it is important to assess the morphing of quasi-symmetric binary fission first into asymmetric binary Coulomb fragmentation, and then into multiple fragmentation. Such an effort would also advance thermodynamic theory of small open quantal systems.

\begin{acknowledgments}
This work was supported by the U.S. Department of Energy grant No.
DE-FG02-88ER40414.
\end{acknowledgments}

\end{document}